\documentclass[twocolumn,tighten]{aastex62}

\submitjournal{ApJ}

\newcommand{\lya}{Ly$\alpha$}
\newcommand{\lyb}{Ly$\beta$}

\newcommand{\hi}{\ion{H}{1}}

\newcommand{\heii}{\ion{He}{2}}

\newcommand{\taueff}{$\tau_{\rm eff}$}

\newcommand{\hinvMpc}{$h^{-1}\,$Mpc}
\newcommand{\rband}{{\it r2}}
\newcommand{\iband}{{\it i2}}
\newcommand{\nb}{{\it NB816}}

\shorttitle{Evidence for UVB Fluctuations}
\shortauthors{Becker et al.}

\begin{document}

\title{Evidence for Large-Scale Fluctuations in the Metagalactic Ionizing Background Near Redshift Six\footnote{Based in part on data collected at the Subaru Telescope, which is operated by the National Astronomical Observatory of Japan.}}

\correspondingauthor{George Becker}
\email{george.becker@ucr.edu}

\author[0000-0003-2344-263X]{George D. Becker}
\affiliation{Department of Physics \& Astronomy, University of California, Riverside, CA, 92521, USA}

\author{Frederick B. Davies}
\affiliation{Department of Physics, University of California, Santa Barbara, CA 93106, USA}

\author{Steven R. Furlanetto}
\affiliation{Department of Physics \& Astronomy, University of California, Los Angeles, CA 90095, USA}

\author{Matthew A. Malkan}
\affiliation{Department of Physics \& Astronomy, University of California, Los Angeles, CA 90095, USA}

\author{Elisa Boera}
\affiliation{Department of Physics \& Astronomy, University of California, Riverside, CA, 92521, USA}

\author{Craig Douglass}
\affiliation{Department of Physics \& Astronomy, University of California, Riverside, CA, 92521, USA}

\begin{abstract}

The observed scatter in intergalactic \lya\ opacity at $z \lesssim 6$ requires large-scale fluctuations in the neutral fraction of the intergalactic medium (IGM) after the expected end of reionization.  Post-reionization models that explain this scatter invoke fluctuations in either the ionizing ultraviolet background (UVB) or IGM temperature.  These models make very different predictions, however, for the relationship between \lya\ opacity and local density.  Here we test these models using \lya-emitting galaxies (LAEs) to trace the density field surrounding the longest and most opaque known \lya\ trough at $z < 6$.  Using deep Subaru Hyper Suprime-Cam narrow-band imaging, we find a highly significant deficit of $z \simeq 5.7$ LAEs within 20 \hinvMpc\ of the trough.  The results are consistent with a model in which the scatter in \lya\ opacity near $z \sim 6$ is driven by large-scale UVB fluctuations, and disfavor a scenario in which the scatter is primarily driven by variations in IGM temperature.  UVB fluctuations at this epoch present a  boundary condition for reionization models, and may help shed light on the nature of the ionizing sources.

\end{abstract}


\section{Introduction} \label{sec:intro}

\begin{figure*}
\plotone{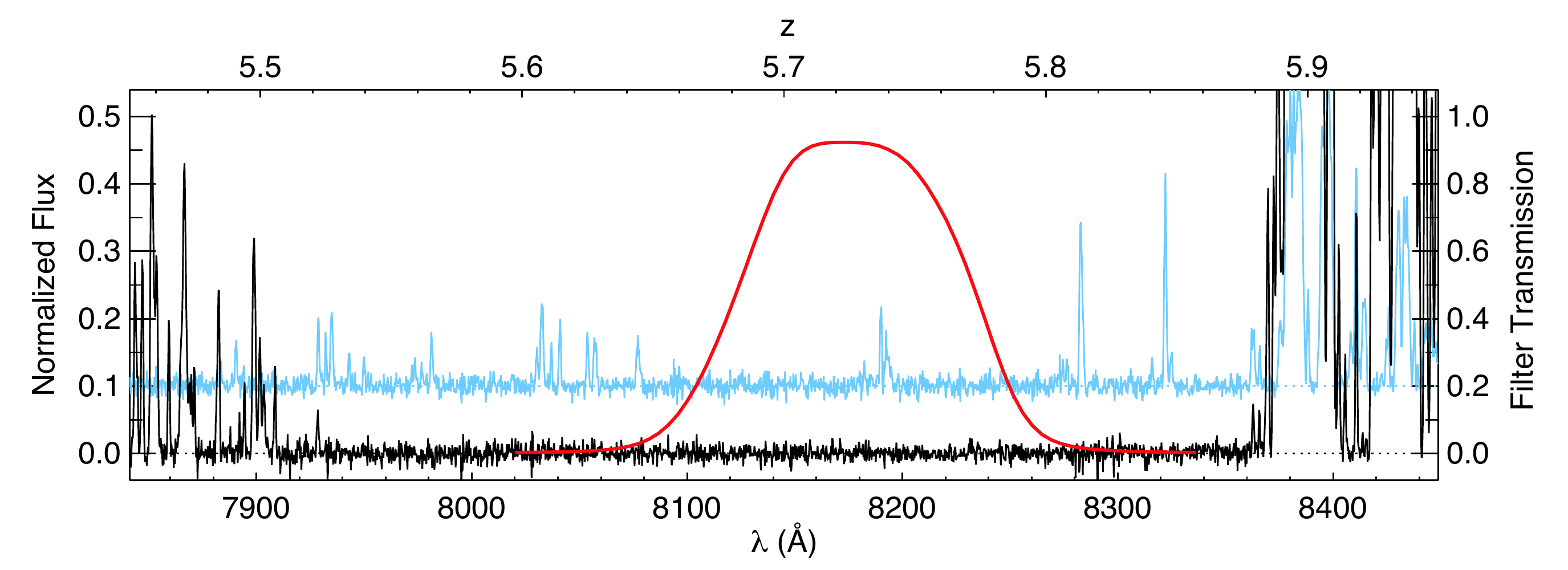}
\caption{X-Shooter spectrum of the $z=6.0$ quasar ULAS J0148+0600, from \citet{becker2015}.  The black line is centered on the \lya\ forest and includes the $\sim$110 \hinvMpc\ \lya\ absorption trough spanning $7930~{\rm \AA} < \lambda < 8360~{\rm \AA}$.  Corresponding redshifts are shown along the top axis.  The blue line, offset by 0.1 in normalized flux, is shifted in wavelength to show the \lyb\ forest at the same redshifts.  The red line shows the HSC \nb\ filter curve.\label{fig:trough}}
\vspace{0.1in}
\end{figure*}

The characteristics of the high-redshift intergalactic medium (IGM) are a key diagnostic of cosmic reionization and the growth of the first galaxies.  The temperature, ionization state, and density distribution of the IGM at $z \gtrsim 5$ reflect the impact of galaxies on their environments at early times.  Spatial fluctuations in IGM properties, moreover, may reflect the late-time impact of inhomogeneous reionization.

One of the most striking features of the IGM at these redshifts is the presence of large-scale fluctuations in \lya\ opacity.  \citet{fan2006b} noted these fluctuations in measurements of the effective optical depth ($\tau_{\rm eff} = -\ln{\langle \mathcal T \rangle}$, where ${\mathcal T}$ is the continuum-normalized transmission) in the \lya\ forest towards 19 $z \sim 6$ quasars \citep[see also][]{djorgovski2006}.  The scatter was confirmed in further measurements by \citet{becker2015} and \citet{bosman2018}, with the most extreme example being a giant ``Gunn-Peterson'' trough spanning 110 \hinvMpc\ towards the $z=6.0$ quasar ULAS J0148+0600 (herein J0148).  Although the scatter in opacity due to variations in the density field can become large when the mean opacity is high \citep{lidz2006}, the range in \lya\ opacity on 50 $h^{-1}$ comoving Mpc scales near $z \sim 6$ significantly exceeds the scatter expected due to the density field alone \citep{becker2015}.  Large-scale, order unity variations in the hydrogen neutral fraction must therefore be present at these redshifts, in stark contrast to the roughly uniform neutral fraction (averaged over large scales) generally assumed for the IGM at later times.

In a photoionized IGM, the hydrogen neutral fraction, $f_{\rm H\, I}$, scales as
\begin{equation}
   f_{\rm H\, I} \propto n_{\rm H}T^{-0.7}\Gamma^{-1} \, ,
\end{equation}
where $n_{\rm H}$ is the total hydrogen density, $T$ is the gas temperature (which impacts the recombination rate), and $\Gamma$ is the photoionization rate.  If density fluctuations alone are insufficient to produce the observed range in \lya\ opacity, then large-scale variations in temperature and/or photoionization rate must be present.  Over the past few years, multiple models have invoked such fluctuations to explain the wide distribution of \lya\ opacities near $z \sim 6$.  \citet{davies2016} proposed that fluctuations in a galaxy-dominated ionizing ultraviolet background (UVB) may be present due to spatial variations in the mean free path of ionizing photons.  \citet{chardin2015,chardin2017} also proposed that the wide \taueff\ distribution may be due to UVB fluctuations, but attributed the fluctuations to a radiation field dominated by rare, bright sources such as quasars.  On the temperature side, \citet{daloisio2015} proposed that large temperature fluctuations may be present following an extended reionization epoch that ended not long before $z = 6$.  

Intriguingly, each of these models poses challenges for conventional IGM models.  In the \citet{davies2016} UVB model, the typical mean free path must be at least a factor of three shorter than what would be predicted from extrapolations of lower-redshift measurements \citep[][and references therein]{worseck2014}.  The evolution of the global ionizing emissivity may also be  unphysically rapid over $5 < z < 6$, unless estimates at $z \sim 5$ are too low due to biases in the measured mean free path \citep{daloisio2018}.  The \citet{chardin2015,chardin2017} model requires a number density of quasars at the high end of observational constraints \citep{giallongo2015,mcgreer2018}.  A UVB dominated by quasars may also cause helium in the IGM to fully reionize too early \citep{daloisio2017}.  This could violate evidence from the \heii\ \lya\ forest that helium reionization ends near $z \sim 3$ \citep[e.g.,][]{worseck2011b}, and produce IGM temperatures that exceed current constraints near $z \sim 4-5$ \citep{becker2011a}.  Finally, the temperature model requires both an extended, late reionization history and a local temperature boost from reionization that is at the upper end of physically motivated values \citep{mcquinn2012}.  It is uncertain, moreover, whether sufficient temperature fluctuations can be produced in radiative transfer simulations of reionization that are consistent with IGM temperature measurements at $z < 5$ (\citealt{keating2017}, but see \citealt{uptonsanderbeck2016}).  Clearly, determining the origin of the \taueff\ fluctuations would shed new light on the physics governing the high-redshift IGM.

Recently, \citet{davies2017} showed that the competing models could be tested by probing the relationship between \lya\ opacity and local environment \citep[see also][]{daloisio2018}.  Specifically, the galaxy UVB model of \citet{davies2016} predicts that a deep \lya\ trough such as the one towards J0148 should arise in voids, where the UV background is suppressed.  The temperature model of \citet{daloisio2015}, in contrast, predicts that troughs should occur in high density regions that reionized early and have had sufficient time to cool.  \citet{davies2017} suggested that galaxies along the quasar line of sight could be used to probe the density field.  Observationally, this test requires (i) identifying galaxies within the redshift range of the trough, (ii) sufficient sensitivity that the galaxies will adequately sample the underlying density field, and (iii) a survey area that is large enough to cover the region of interest around the quasar line of sight and, ideally, a surrounding region that can be used for a self-consistent comparison.

In this paper we report on a survey for \lya\ emitting galaxies (LAEs) in the field of J0148 using Hyper Suprime-Cam (HSC) on the Subaru telescope.  The HSC data have sufficient areal coverage and depth to conduct the experiment proposed by \citet{davies2017}.  In addition, the LAE candidates are selected using the NB816 narrow-band filter, whose central wavelength conveniently sits right in the middle of the J0148 trough (Figure~\ref{fig:trough}).  We present our HSC data in Section~\ref{sec:data}.  The selection of LAE candidates is described in Section~\ref{sec:laes}, and the results are compared to model predictions in Section~\ref{sec:models}.  Finally, we summarize our results and discuss caveats to the interpretation in Section~\ref{sec:summary}.  Throughout the paper we assume a $\Lambda$CDM cosmology with $\Omega_m = 0.3$, $\Omega_\Lambda = 0.7$, $\Omega_{b} = 0.048$, and $\sigma_8 = 0.82$, and quote distances in comoving units.

\section{Data} \label{sec:data}

\subsection{Hyper Suprime-Cam Imaging}\label{sec:hsc}

We acquired deep HSC imaging of the field around J0184 in September 2016 and August 2017, with the majority of data obtained in 2017.  Imaging was performed in the \rband, \iband, and \nb\ filters in dithered exposures centered on the quasar position.  Total exposure times are listed in Table~\ref{tab:imaging}.  The \nb\ filter has a transmission-weighted mean wavelength of $\lambda = 8177$~\AA, corresponding to \lya\ at $z = 5.726$, and $\ge$50\% peak transmission over  $8122~{\rm \AA} < \lambda < 8239~{\rm \AA}$, corresponding to $5.681 < z < 5.777$.

The data were processed using the LSST Science Pipeline\footnote{\url https://pipelines.lsst.io} Version 13, release w\_2017\_28 \citep{axelrod2010,juric2015}, with an implementation closely following the HSC Software Pipeline described in \citet{bosch2018}.  The pipeline processes the individual CCDs and combines them into stacked mosaics.  PanSTARRS1 DR1 imaging \citep{chambers2016} was used for photometric and astrometric calibration.  The pipeline also estimates the seeing parameters as a function of position.  The median seeing in all bands was roughly 0.6'' (Table~\ref{tab:imaging}).
 
 The LSST pipeline performs a range of photometric measurements.  For our analysis we used {\tt forced} measurements, wherein the intrinsic spatial parameters of a source are determined from the band in which it is detected with the highest significance (typically \nb\ for our sources).  These parameters are then held fixed for other bands, and the model profile is convolved with the local point spread function when measuring fluxes.  This approach is useful for determining accurate colors of extended sources.  Except where noted below, we adopt CModel magnitudes, which use a composite of the best fit exponential and de Vaucouleurs profiles \citep{abazajian2004,bosch2018}.  The median 5$\sigma$ limiting CModel magnitudes are listed in Table~\ref{tab:imaging}.  For reference, we also include 5$\sigma$ limits for PSF magnitudes and for magnitudes measured within 1.5 arcsecond apertures.  The values in Table~\ref{tab:imaging} are the magnitudes at which 50\% of all detected sources have signal-to-noise ratios $S/N \ge 5$.  We verified that, for the circular aperture and PSF cases, these values are very similar (within 0.05 magnitudes) to the limits directly measured at randomly placed positions within 40 arcminutes of the field center that are free of sources.  In all bands the sensitivity declines by ~0.2 magnitudes from the center of the field to a radius of 40 arcminutes, then deteriorates rapidly at larger radii.  The majority of our analysis is therefore confined to the inner 40 arcminutes.

\subsection{J0148 \lya\ Opacity}

Before proceeding to the LAE selection, we note that our imaging data allows us to check the \lya\ \taueff\ measurement along the J0148 line of sight.  Here we use PSF fluxes, since the object is known to be point-like.  We measure a narrow-band flux from the quasar of $F_{\lambda}^{NB816} = (1.9 \pm 1.0) \times 10^{-20}~{\rm erg\, s^{-1}\,cm^{-2}\,\AA}$.  We meanwhile measure an \iband\ flux from the quasar of $F_{\lambda}^{i2} = 3.7 \times 10^{-18}~{\rm erg\, s^{-1}\,cm^{-2}\,\AA}$, with negligible error.  We use the \iband\ flux to scale the X-Shooter spectrum presented in \citet{becker2015} by convolving the spectrum with the \iband\ transmission curve, and estimate an unabsorbed continuum flux at the \nb\ wavelength of $F_{\lambda}^{\rm cont} = 2.0 \times 10^{-17}~{\rm erg\, s^{-1}\,cm^{-2}\,\AA}$.  These measurements translate into an effective optical depth over the \nb\ wavelength range of $\tau_{\rm eff} = -\ln{(F_{\lambda}^{NB816}/F_{\lambda}^{\rm cont})} = 6.94^{+0.68}_{-0.40}$ (1$\sigma$), with a 2$\sigma$ lower limit of $\tau_{\rm eff} \ge 6.26$
\footnote{In spectroscopic measurements where $F \le 2\sigma_{F}$, it is conventional to report a lower limit on the effective optical depth of $\tau_{\rm eff} \ge -\ln{(2\sigma_{F}/F^{\rm cont})}$, where $\sigma_{F}$ is the uncertainty in the mean flux \citep[e.g.,][]{fan2006b}.  By this convention, the photometric limit measured here would be $\tau_{\rm eff} \ge 6.9$.}  These values are consistent with the $2\sigma$ limit of $\tau_{\rm eff} \ge 7.2$ measured by \citet{becker2015} for the 50~\hinvMpc\ section centered at $z=5.796$, which largely overlaps but is $\sim$70\% longer than the \nb\ passband.

\begin{deluxetable}{cccccc}
\tablecaption{Imaging Data Summary\label{tab:imaging}}
\tablewidth{0pt}
\tablehead{
\colhead{Filter} & {$t_{\rm exp}$} & \colhead{Seeing\tablenotemark{a}} & 
\multicolumn{3}{c}{$m_{5\sigma}$\tablenotemark{b}} \\
\colhead{} & \colhead{(hr)} & \colhead{(arcsec)} & \colhead{1.5\arcsec} & \colhead{PSF} & \colhead{CModel}
}
\startdata
\rband  &  1.5  &  0.62  &  26.5  &  27.5  &  27.4  \\  
\iband  &  2.4  &  0.64  &  26.0  &  27.0  &  26.9  \\  
\nb     &  4.5  &  0.60  &  25.3  &  26.3  &  26.1 
\enddata
\tablenotetext{a}{Median seeing FWHM across all source positions in the combined mosaic}
\tablenotetext{b}{Magnitude at which 50\% of detected sources detected have $S/N \ge 5$.  Values are given for fluxes within 1.5\arcsec circular apertures, as well as for PSF and CModel fluxes.} 
\end{deluxetable}

\subsection{LAE Selection}\label{sec:laes}

\begin{figure}
   \centering
   \includegraphics[width=0.42\textwidth]{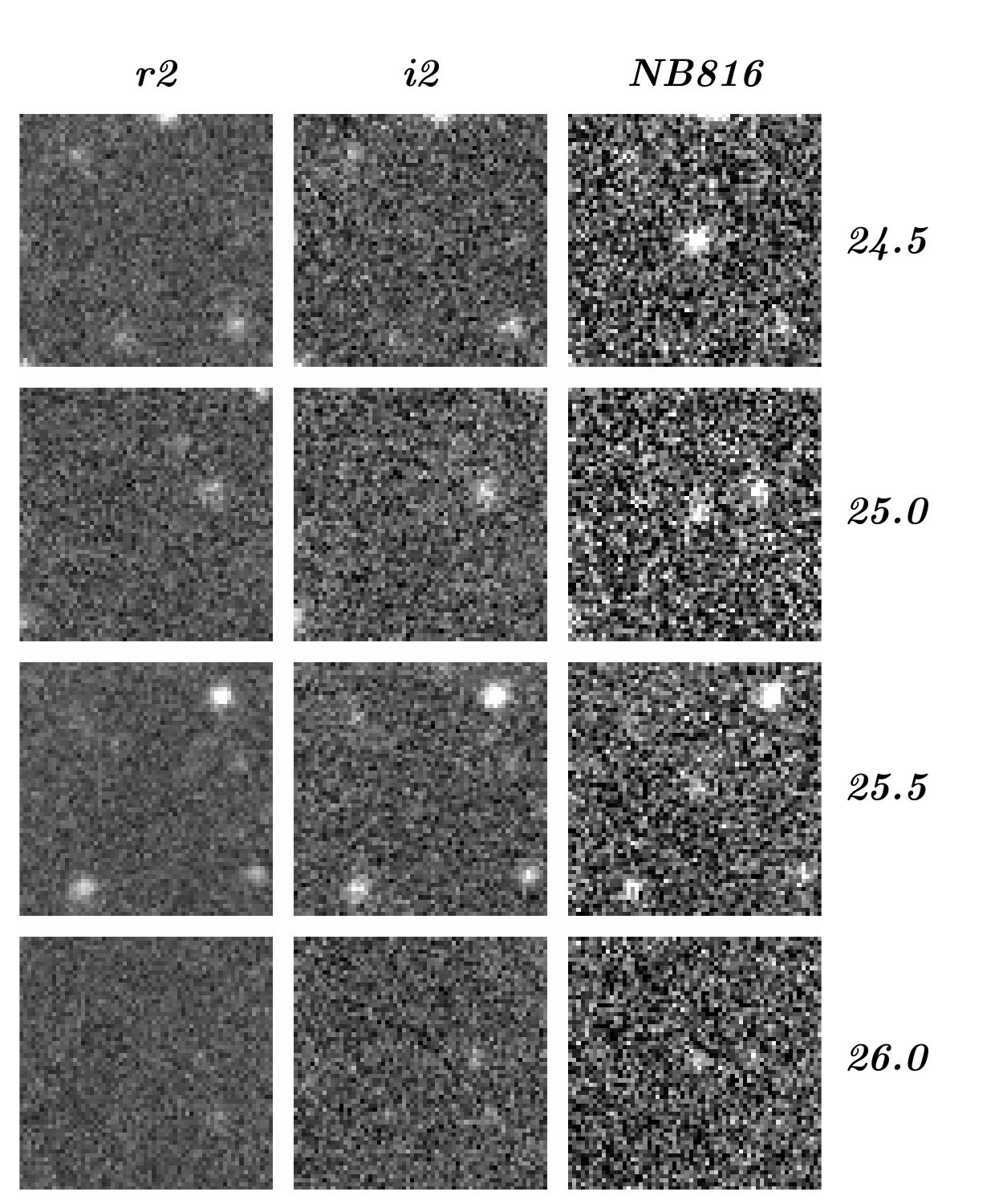}
   \caption{Cutout images for a selection of LAE candidates.  Images are 10\arcsec on a side and are centered on the candidate position.  From left to right, columns show the \rband, \iband, and \nb\ bands.  The \nb\ magnitude for each candidate is displayed on the right.  These objects were chosen to have $S/N$ ratios near the median among objects of similar magnitude.  \label{fig:thumbnails}}
\end{figure}

Candidate $z = 5.7$ LAEs are selected to have a narrow-band excess indicative of \lya\ emission and an $r2-i2$ color consistent with \lya\ forest absorption by the IGM.  Following \citet{ouchi2008}, we impose the following selection criteria:
\begin{eqnarray}
&i2 - NB816 \ge 1.2,~{\rm and}\nonumber \\
&\left[ (r2 > 2\sigma_{r2})~{\rm or}~(r2 \le 2\sigma_{r2}~{\rm and}~ r2-i2 \ge 1.0)   \right]
\end{eqnarray}
Similar to \citet{ono2018} and \citet{shibuya2017} we also select on a number of pipeline flags to avoid blended or contaminated sources.  For a list of flags see \cite{shibuya2017}.  Source are chosen to have $NB816 \le 26.0$ and $S/N(NB816) \ge 5$, and to lie within 45\arcmin of the quasar position.  Each source satisfying these criteria is visually inspected in the combined images.  For those passing this initial visual inspection, individual \nb\ exposures are examined as a further check for moving or spurious sources.  In total, 806 LAE candidates pass our selection process.  Their properties are summarized in Table~\ref{tab:properties}.  Cutout images for four of these objects spanning a range of \nb\ magnitude are shown in Figure~\ref{fig:thumbnails}.

\begin{deluxetable*}{cccccccc}
\tablecaption{Properties of LAE Candidates\label{tab:properties}}
\tablewidth{0pt}
\tablehead{
\colhead{ID} & {$\alpha$} & \colhead{$\delta$} &  \colhead{\nb}  & \colhead{${\mathcal F}(r2)$}  &
  \colhead{${\mathcal F}(i2)$}  &  \colhead{${\mathcal F}(\mbox{\nb})$}  &  \colhead{$\log{L_{\rm Ly\alpha}}$} \\
\colhead{ }  &  \colhead{(J2000)}  &  \colhead{(J2000)}  &  \colhead{(mag)}  &  
  \colhead{(mJy)}  &  \colhead{(mJy)}  &  \colhead{(mJy)}  &  \colhead{erg s$^{-1}$} \\
  \colhead{(1)}  &  \colhead{(2)}  &    \colhead{(3)}  &  \colhead{(4)}  &  \colhead{(5)}  &  \colhead{(6)}  &  \colhead{(7)}  &  \colhead{(8)}
}
\startdata
J014551+055853  &  01 45 51.453  &  +05 58 53.74  &  25.53  &   0.000$\pm$0.016  &   0.036$\pm$0.026  &  0.224$\pm$0.035  &  $42.623_{-0.073}^{0.062}$  \\
J014553+061419  &  01 45 53.708  &  +06 14 19.60  &  25.00  &   0.031$\pm$0.015  &   0.101$\pm$0.024  &  0.362$\pm$0.044  &  $42.832_{-0.057}^{0.050}$  \\
J014554+061920  &  01 45 54.792  &  +06 19 20.72  &  25.61  &  -0.004$\pm$0.010  &  -0.000$\pm$0.016  &  0.207$\pm$0.030  &  $42.590_{-0.067}^{0.058}$  \\
J014556+060639  &  01 45 56.396  &  +06 06 39.90  &  25.89  &   0.012$\pm$0.010  &   0.043$\pm$0.018  &  0.161$\pm$0.028  &  $42.479_{-0.083}^{0.069}$  \\
J014556+055250  &  01 45 56.989  &  +05 52 50.80  &  25.73  &  -0.010$\pm$0.014  &   0.056$\pm$0.017  &  0.186$\pm$0.031  &  $42.542_{-0.078}^{0.066}$  \\
J014557+060459  &  01 45 57.299  &  +06 04 59.24  &  25.04  &   0.001$\pm$0.012  &   0.002$\pm$0.023  &  0.350$\pm$0.039  &  $42.817_{-0.052}^{0.046}$  \\
J014557+061720  &  01 45 57.383  &  +06 17 20.58  &  25.48  &   0.008$\pm$0.011  &   0.027$\pm$0.015  &  0.233$\pm$0.034  &  $42.640_{-0.069}^{0.059}$  \\
J014557+060324  &  01 45 57.545  &  +06 03 24.84  &  24.84  &   0.034$\pm$0.011  &   0.096$\pm$0.020  &  0.423$\pm$0.031  &  $42.899_{-0.033}^{0.031}$  \\
J014557+060606  &  01 45 57.898  &  +06 06 06.41  &  24.97  &   0.024$\pm$0.013  &   0.097$\pm$0.020  &  0.373$\pm$0.033  &  $42.844_{-0.040}^{0.037}$  \\
J014558+054525  &  01 45 58.048  &  +05 45 25.19  &  25.53  &   0.004$\pm$0.011  &   0.020$\pm$0.017  &  0.222$\pm$0.035  &  $42.620_{-0.075}^{0.064}$  \\
J014558+061023  &  01 45 58.477  &  +06 10 23.17  &  25.24  &  -0.024$\pm$0.010  &   0.011$\pm$0.018  &  0.291$\pm$0.030  &  $42.737_{-0.046}^{0.042}$  \\
J014559+054157  &  01 45 59.199  &  +05 41 57.48  &  25.83  &  -0.020$\pm$0.010  &  -0.008$\pm$0.015  &  0.169$\pm$0.028  &  $42.502_{-0.077}^{0.065}$  \\
J014559+061720  &  01 45 59.492  &  +06 17 20.25  &  24.49  &   0.009$\pm$0.014  &   0.046$\pm$0.024  &  0.579$\pm$0.036  &  $43.036_{-0.028}^{0.026}$  \\
J014559+054349  &  01 45 59.540  &  +05 43 49.56  &  25.45  &   0.014$\pm$0.009  &   0.055$\pm$0.013  &  0.241$\pm$0.024  &  $42.655_{-0.046}^{0.042}$  \\
J014600+054322  &  01 46 00.030  &  +05 43 22.60  &  25.49  &  -0.009$\pm$0.012  &   0.023$\pm$0.016  &  0.231$\pm$0.031  &  $42.637_{-0.062}^{0.055}$  \\
J014600+061908  &  01 46 00.132  &  +06 19 08.00  &  25.23  &  -0.039$\pm$0.012  &   0.037$\pm$0.021  &  0.295$\pm$0.037  &  $42.743_{-0.059}^{0.052}$  \\
J014600+061505  &  01 46 00.497  &  +06 15 05.02  &  25.46  &   0.023$\pm$0.009  &   0.060$\pm$0.014  &  0.237$\pm$0.026  &  $42.649_{-0.051}^{0.046}$  \\
J014600+054204  &  01 46 00.724  &  +05 42 04.68  &  25.70  &  -0.019$\pm$0.012  &   0.016$\pm$0.017  &  0.191$\pm$0.030  &  $42.555_{-0.075}^{0.064}$  \\
J014600+060622  &  01 46 00.896  &  +06 06 22.00  &  25.90  &   0.000$\pm$0.008  &   0.030$\pm$0.015  &  0.158$\pm$0.024  &  $42.472_{-0.073}^{0.063}$  \\
J014601+061112  &  01 46 01.607  &  +06 11 12.99  &  25.82  &  -0.011$\pm$0.008  &   0.022$\pm$0.014  &  0.170$\pm$0.026  &  $42.504_{-0.072}^{0.062}$ 
\enddata
\tablecomments{(1) Identifier; (2) Right ascension; (3) Declination; (4) \nb\ magnitude; (5) \rband\ flux; (6) \iband\ flux; (7) \nb\ flux; (8) Log \lya\ luminosity.  A complete list of sources is available in the online version of this table.}
\end{deluxetable*}

\begin{figure}
   \centering
   \includegraphics[width=0.47\textwidth]{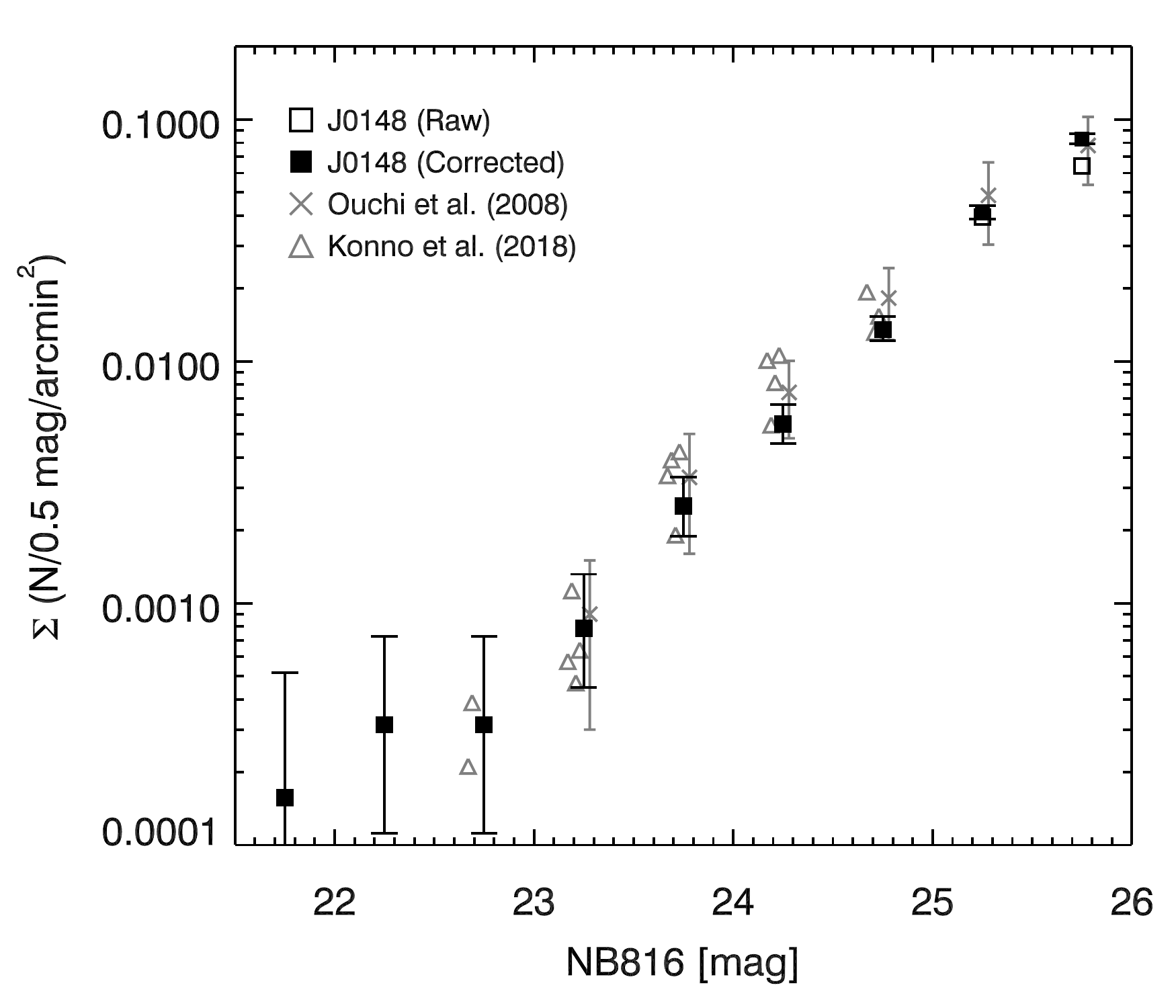}
   \caption{Surface density of LAE candidates as a function of \nb\ magnitude.  The open squares show our raw results averaged over the entire HSC field.  Filled squares have been approximately corrected for completeness.  Note that the raw and corrected values are only significantly different in the faintest bin.  Error bars for the J0148 field are 68\% Poisson intervals.  Completeness-corrected values from \citet{ouchi2008} and \citet{konno2018} are shown for comparison.  Values for the four HSC fields of \citet{konno2018} are plotted separately, and include only magnitude bins where the completeness correction is moderate (up to $\sim$tens of per cent).\label{fig:sigma_compare}}
\end{figure}

The goal of this paper is to determine the relative density of LAEs near the quasar line of sight by self-consistently comparing this region to the outskirts of the HSC field.  We therefore do not attempt  to make a detailed completeness estimate that would be needed to determine a luminosity function.  Nevertheless, we can compare our number counts to previous results from the literature.  In Figure~\ref{fig:sigma_compare} we plot the surface density of LAE candidates in the J0148 field as a function of \nb\ magnitude.  Raw results are shown along with results roughly corrected for completeness, where the completeness within each half magnitude bin is estimated as the fraction of all \nb\ sources in that bin that meet our $S/N \ge 5$ cut.  For comparison we plot the surface density of LAEs from the Subaru Suprime-Cam observations of \citet{ouchi2008}, and the shallower Hyper Suprime-Cam observations of \citet{konno2018} \citep[previously described in][]{shibuya2017}, both corrected for  completeness.  Overall we find good agreement with the surface densities of these works, which gives us confidence in our selection process.  We note that some level of contamination is expected in our sample.  We address this in Section~\ref{sec:caveats}.

\begin{figure*}
   \centering
   \begin{minipage}{\textwidth}
   \begin{center}
   \includegraphics[width=0.7\textwidth]{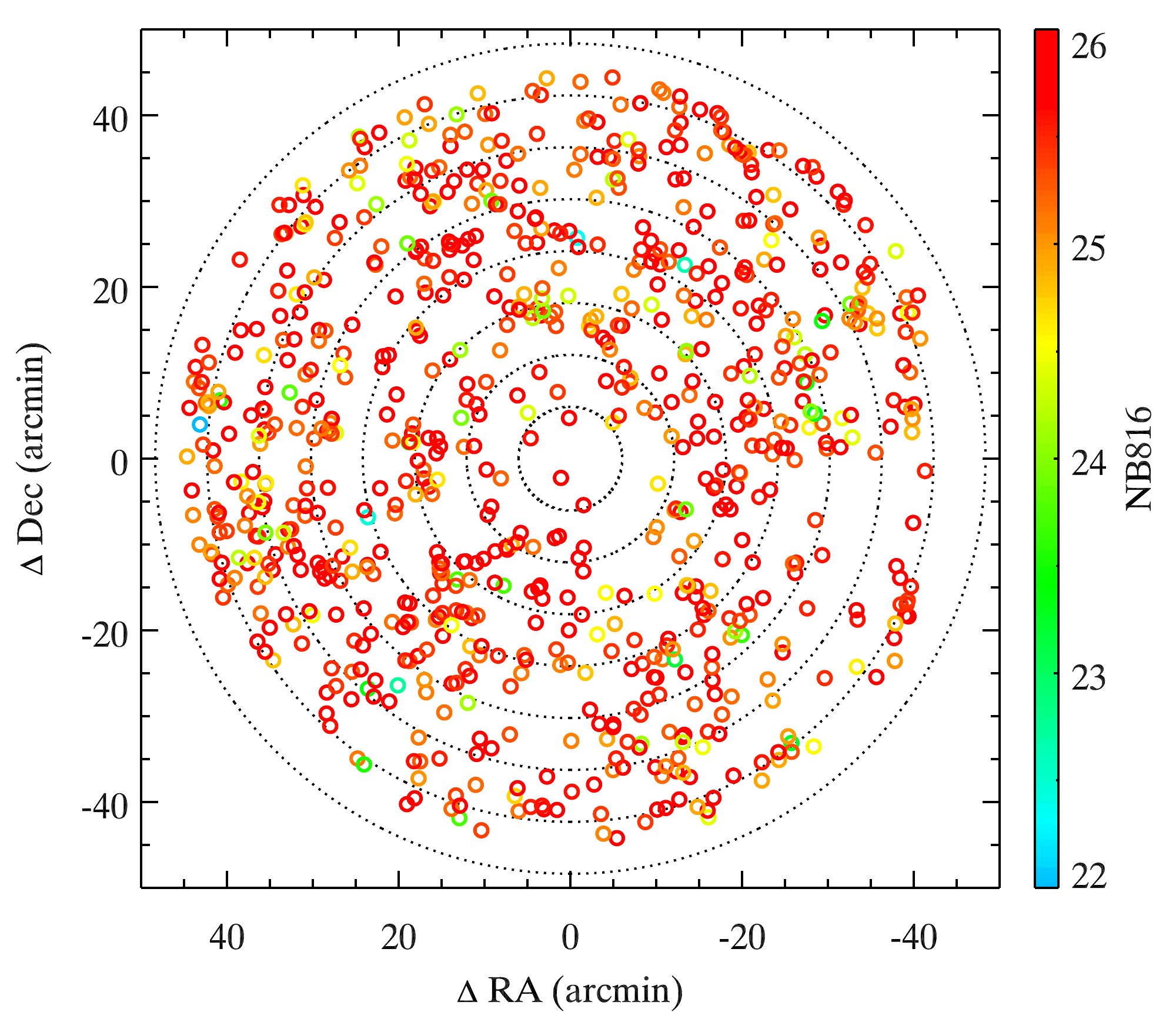}
   \caption{Distribution of LAE candidates on the sky.  The field is centered on the quasar position.  North is up and East is to the left.  Each symbol represents an LAE candidate, and is color-coded by \nb\ magnitude according to the scale on the right.  Dotted lines show concentric circles in increments of 10~\hinvMpc\ projected distance from the quasar line of sight.\label{fig:sky}}
   \end{center}
   \end{minipage}
   \vspace{0.1in}
\end{figure*}

\section{Results}\label{sec:results}

\subsection{Spatial distribution of LAE candidates}

The distribution of LAE candidates on the sky is shown in Figure~\ref{fig:sky}.  Symbols marking the location of LAE candidates are color-coded according to their \nb\ magnitude.  Dotted circles centered on the quasar position are drawn in increments of 10 \hinvMpc\ projected distance.  As noted above, our sensitivity is roughly constant out to $\Delta \theta \simeq 40$\arcmin\ ($R \simeq 66$ \hinvMpc).    

\begin{figure}
   \includegraphics[width=0.48\textwidth]{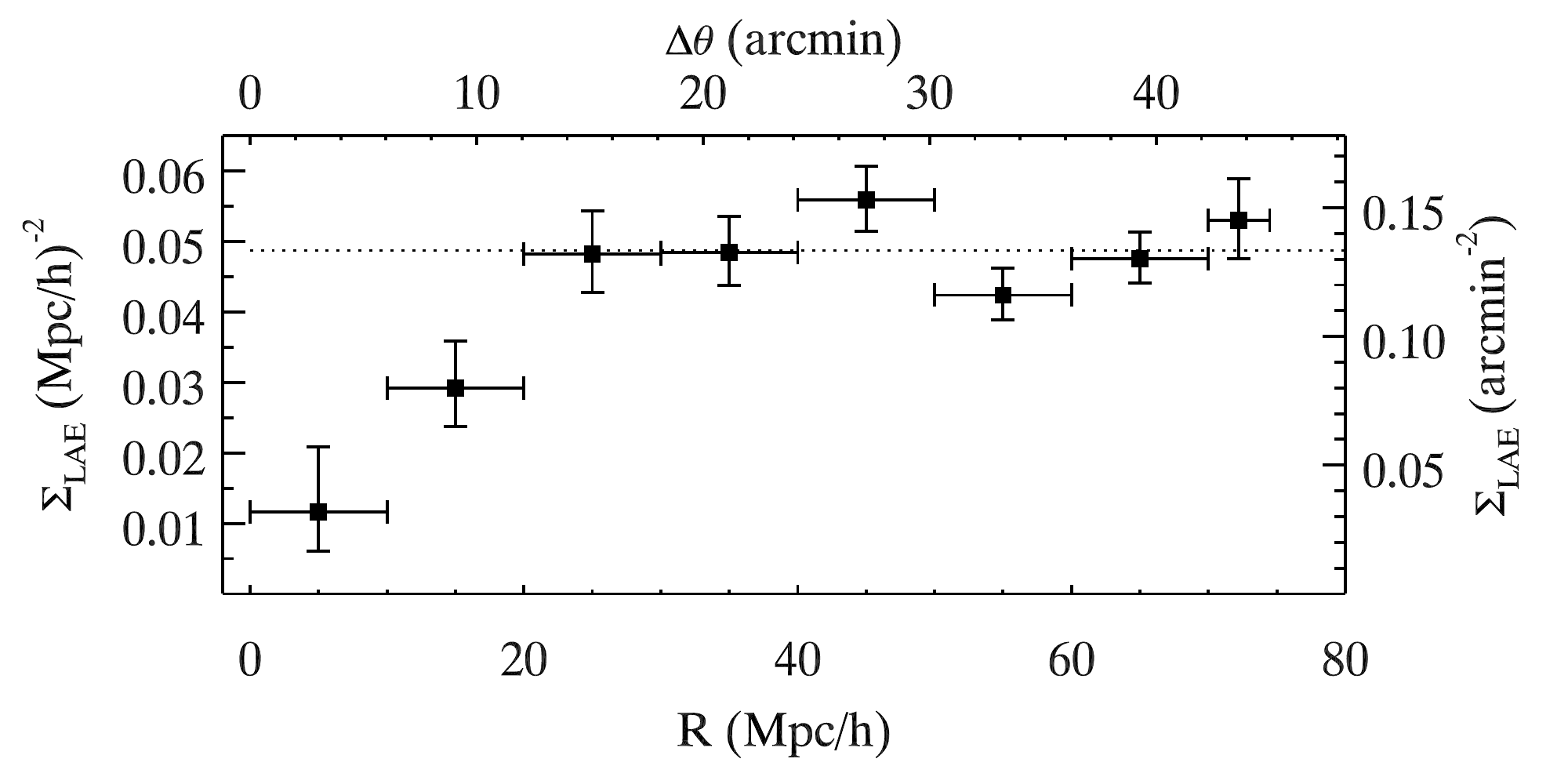}
   \caption{Surface density of LAE candidates as a function of projected distance from the quasar position averaged over annular bins.  Aperture widths are 10~\hinvMpc\ for all except the outermost bin, for which it is 4.5~\hinvMpc.  Vertical error bars are 68\% Poisson intervals (see Table~\ref{tab:results}).  The dotted line shows the mean surface density over $15\arcmin \le \Delta\theta \le 40\arcmin$ from the quasar position.\label{fig:profile}}
\end{figure}

\begin{figure*}
   \centering
   \begin{minipage}{\textwidth}
   \begin{center}
   \includegraphics[width=0.7\textwidth]{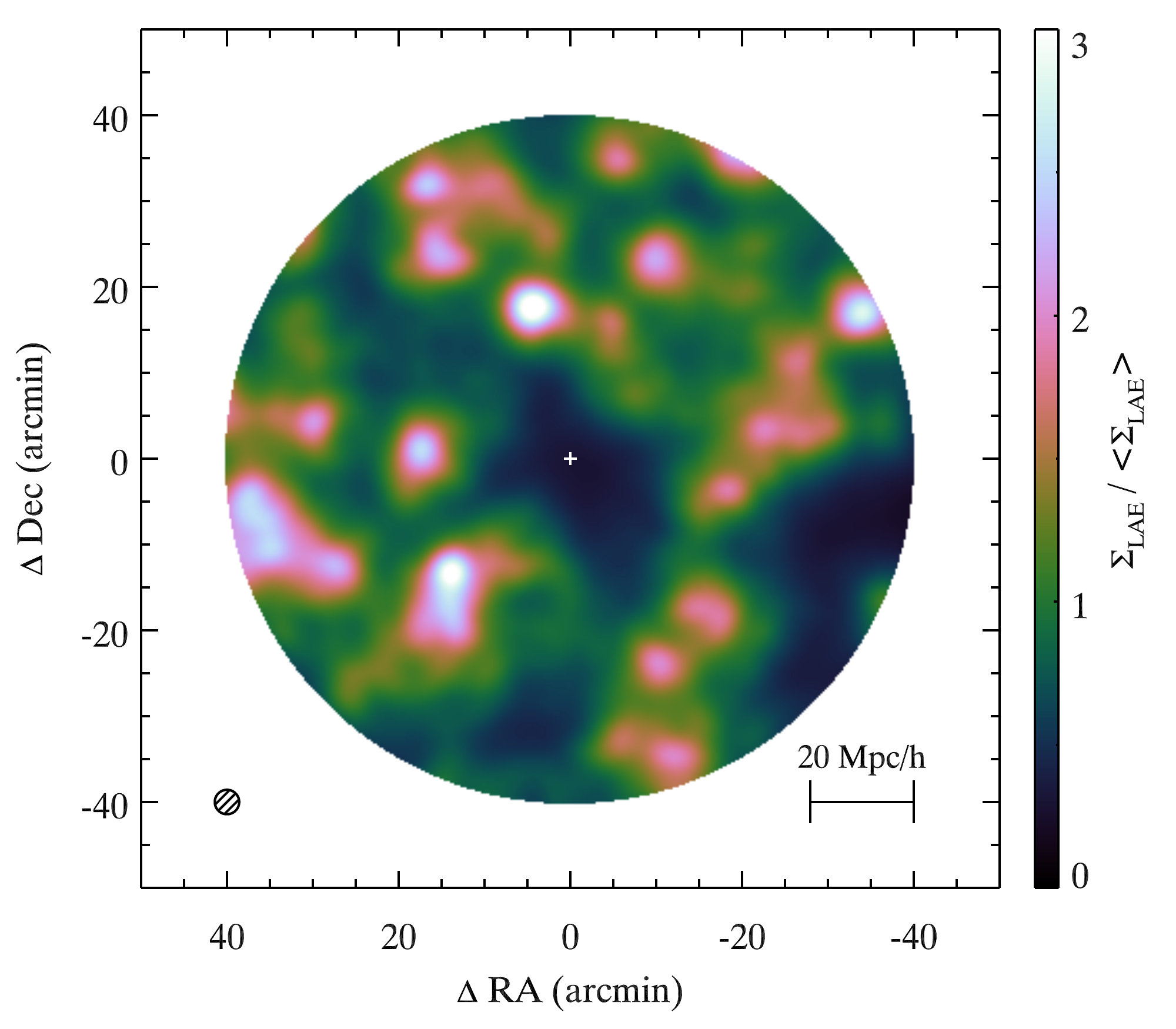}
   \caption{Surface density map of LAE candidates.  The surface density at a given position is calculated from the distance to the tenth nearest source as $\Sigma_{\rm LAE} \propto d_{10}^{-2}$.  Surface densities are displayed as a fraction of the mean value over the field according to the color bar at the right.  The quasar position is marked with a cross.  North is up and East is to the left.  The small circle at lower left shows the FWHM of the smoothing kernel applied to the map.\label{fig:map}}
   \end{center}
   \end{minipage}
   \vspace{0.1in}
\end{figure*}

We compute the surface density of LAEs in radial bins centered on the quasar position.   We use 10 \hinvMpc\ bins in all but the outermost annulus, which is truncated at 45\arcmin\ (74.5 \hinvMpc).  Raw surface densities are corrected by a factor $\Sigma_{NB816}^{\rm all}/\Sigma_{NB816}^{R}$, where $\Sigma_{NB816}^{R}$ is the number density of all narrow-band sources (not just LAE candidates) with $NB816 \le 26$ and $S/N \ge 5$ within a radial bin, and $\Sigma_{NB816}^{\rm all}$ is the number density of narrow-band sources meeting these criteria within $\Delta \theta \le 40$\arcmin.  These corrections, which are less than 10\%, are meant to account for radial variations in sensitivity, as well as for missing areal coverage due to the presence of bright sources.  Our results are summarized in Table~\ref{tab:results}, and the corrected surface density of LAE candidates is plotted as a function of radius in Figure~\ref{fig:profile}.  The dotted line shows the mean ``background'' surface density computed over  $15\arcmin \le \Delta\theta \le 40\arcmin$, which is 0.049~(\hinvMpc)$^{-2}$.  The number density of LAE candidates is consistent with this value in all bins at $R \ge 20$~\hinvMpc.  Near the quasar position, however, the number density is significantly lower.  We find only one candidate within 5 \hinvMpc\ (3.0\arcmin) of the quasar line of sight, and four within 10 \hinvMpc\ (6.0\arcmin), or roughly one quarter of the background density.

We can assess the significance of the underdensity of LAE candidates around the quasar in two ways.  In terms of Poisson statistics, for a background number density of 0.049~(\hinvMpc)$^{-2}$, the expected number of detections within 10 (20) \hinvMpc\ is 15.3 (61.2), and the probability of detecting 4 (28) or fewer candidates is $7 \times 10^{-4}$ ($3 \times 10^{-5}$).  These values ignore the somewhat higher sensitivity at the center of the field, but they also ignore clustering.  For randomly placed circular apertures of radius 10 (20) \hinvMpc\ that fall entirely within 40\arcmin\ of the quasar position, the fraction containing at most 4 (32) candidates is 0.037 (0.027).  The low density of LAE candidates within 20 \hinvMpc\ of the line of sight towards J0148 therefore appears to be highly significant.

To better visualize the spatial distribution of LAE candidates we create a source density map.  To do this, we superimpose a regular grid of 0.4~\hinvMpc (0.24\arcmin) pixels onto the map of LAE candidates shown in Figure~\ref{fig:sky}.  At each grid point, we find the distance, $d_{10}$, to the tenth nearest candidate.  This distance is converted into a source surface density as $\Sigma_{\rm LAE} \propto d_{10}^{-2}$.  The $\Sigma_{\rm LAE}$ values are normalized by their mean value, and the grid is smoothed using a Gaussian kernel of $\sigma = 2.0$~\hinvMpc.  Finally, we truncate the map at $\Delta \theta = 40\arcmin$ to exclude regions with lower sensitivity.  The result is shown in Figure~\ref{fig:map}.  The quasar line of sight passes near the center of an elongated low-density region that extends roughly 40~\hinvMpc\ North-South and 20~\hinvMpc\ East-West.  This is another demonstration that the extreme Gunn-Peterson trough towards J0148 is associated with a region that is highly underdense in LAEs.

\subsection{Comparison to Models}\label{sec:models}

We now turn to comparing our results to predictions from models for the large spread in IGM \lya\ opacities near $z \sim 6$.  Here we consider the UVB fluctuation model of \citet{davies2016} and the temperature fluctuation model of \citet{daloisio2015}, in which deep Gunn-Peterson troughs such as the one towards J0148 arise in under- and over-dense regions, respectively.  Below we briefly describe the implementation of these models in \citet{davies2017} and their predictions for LAEs in the J0148 Gunn-Peterson trough.  We also consider a model in which strong UVB fluctuations are driven by rare, bright sources (QSOs), as in \citet{chardin2015,chardin2017}. 

\begin{deluxetable*}{ccccc}
\tablecaption{LAE Number Density\label{tab:results}}
\tablewidth{0pt}
\tablehead{
\colhead{$R$} & \colhead{$\Delta\theta$} & \colhead{$N_{\rm LAE}$} & \colhead{Correction} & \colhead{$\Sigma_{\rm LAE}$} \\
\colhead{(\hinvMpc)} & \colhead{(arcmin)} & \colhead{} & \colhead{} & \colhead{(\hinvMpc)$^{-2}$}
}
\startdata
   5.0  (0.0--10.0)  &   3.0  (0.0--6.0)   &    4   &  0.91  &  0.012 (0.006--0.021)  \\
  15.0 (10.0--20.0)  &   9.1  (6.0--12.1)  &   28   &  0.98  &  0.029 (0.024--0.036)  \\
  25.0 (20.0--30.0)  &  15.1 (12.1--18.1)  &   77   &  0.98  &  0.048 (0.043--0.054)  \\
  35.0 (30.0--40.0)  &  21.2 (18.1--24.2)  &  107   &  0.99  &  0.048 (0.044--0.054)  \\
  45.0 (40.0--50.0)  &  27.2 (24.2--30.2)  &  156   &  1.01  &  0.056 (0.051--0.061)  \\
  55.0 (50.0--60.0)  &  33.2 (30.2--36.3)  &  146   &  1.00  &  0.042 (0.039--0.046)  \\
  65.0 (60.0--70.0)  &  39.3 (36.3--42.3)  &  191   &  1.02  &  0.048 (0.044--0.051)  \\
  72.2 (70.0--74.5)  &  43.7 (42.3--45.0)  &   97   &  1.10  &  0.053 (0.048--0.059)
\enddata
\tablecomments{Values in parentheses for $R$ and $\Delta\theta$ give the range subtended by each annual bin.  Surface densities are computed as $\Sigma_{\rm LAE} = {\rm Correction} \times N_{\rm LAE} / {\rm Area}$.  The correction factor roughly accounts for sensitivity variations and loss of areal coverage due to bright sources.  Values in parentheses for $\Sigma_{\rm LAE}$ are corrected 68\% Poisson intervals.}
\end{deluxetable*}

\begin{figure*}
   \centering
   \begin{minipage}{\textwidth}
   \begin{center}
   \includegraphics[width=0.7\textwidth]{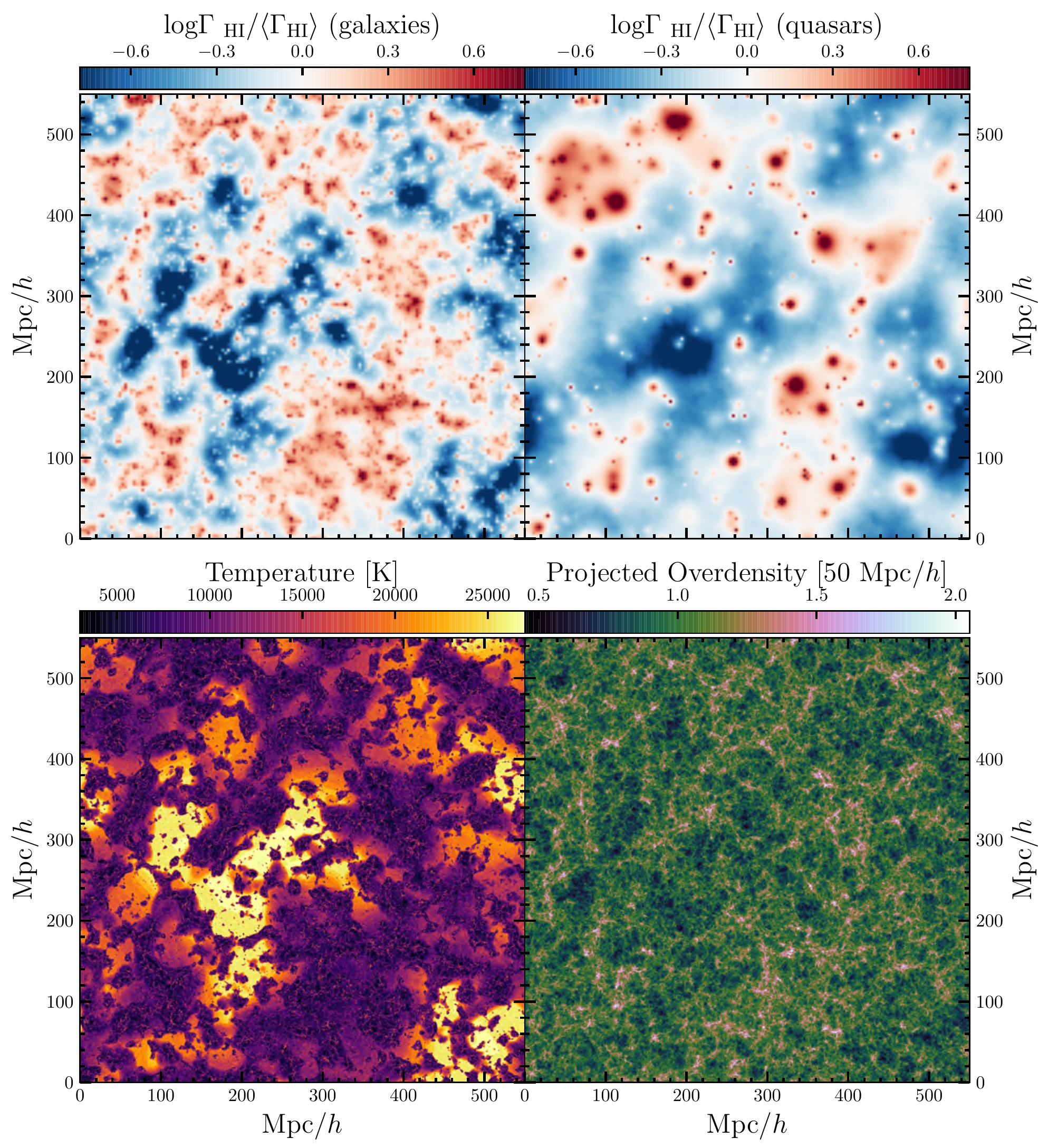}
   \caption{Top left: A 5 \hinvMpc-thick slice through the ionizing radiation field in the galaxy UVB model, where the \hi\ ionization rate, $\Gamma_{\rm H\, I}$ is displayed as a fraction of its mean value.  Top right: Same, but for the QSO UVB model.  Bottom left: A 400 $h^{-1}$ kpc-thick slice through the temperature fluctuation model.  Bottom right: The density field from which these models were generated.  Note the tight correlation between density and $\Gamma_{\rm H\, I}$ in the galaxy UVB model, and the somewhat weaker correlation in the QSO UVB model.  All panels except the top right are adapted from \citet{davies2017}.\label{fig:model_maps}}
   \end{center}
   \end{minipage}
   \vspace{0.1in}
\end{figure*}

\begin{figure}
   \centering
   \includegraphics[width=0.42\textwidth]{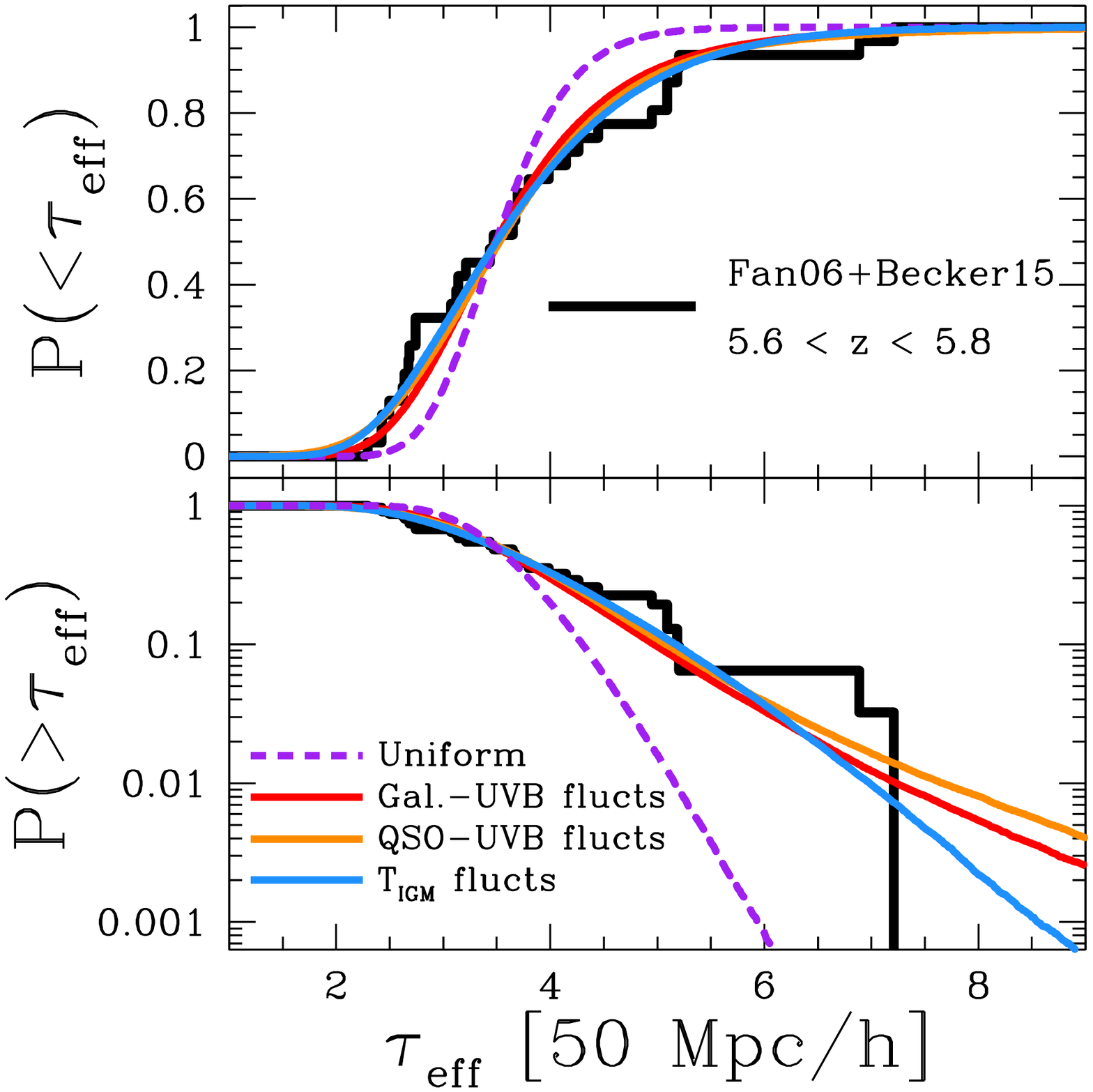}
   \caption{Cumulative distribution of \lya\ effective optical depths measured on 50~\hinvMpc\ scales.  Histograms are data from \citet{fan2006b} and \citet{becker2015}.  Dashed lines are for a model with a uniform UVB and temperature-density relation.  Solid lines are for the galaxy UVB, QSO UVB, and temperature fluctuation models.  The uniform, galaxy UVB, and temperature models are adapted from \citet{davies2017}.  The quasar UVB model is described in Section~\ref{sec:models}.\label{fig:taudist}}
\end{figure}

\begin{figure*}
   \centering
   \begin{minipage}{\textwidth}
   \begin{center}
   \includegraphics[width=0.95\textwidth]{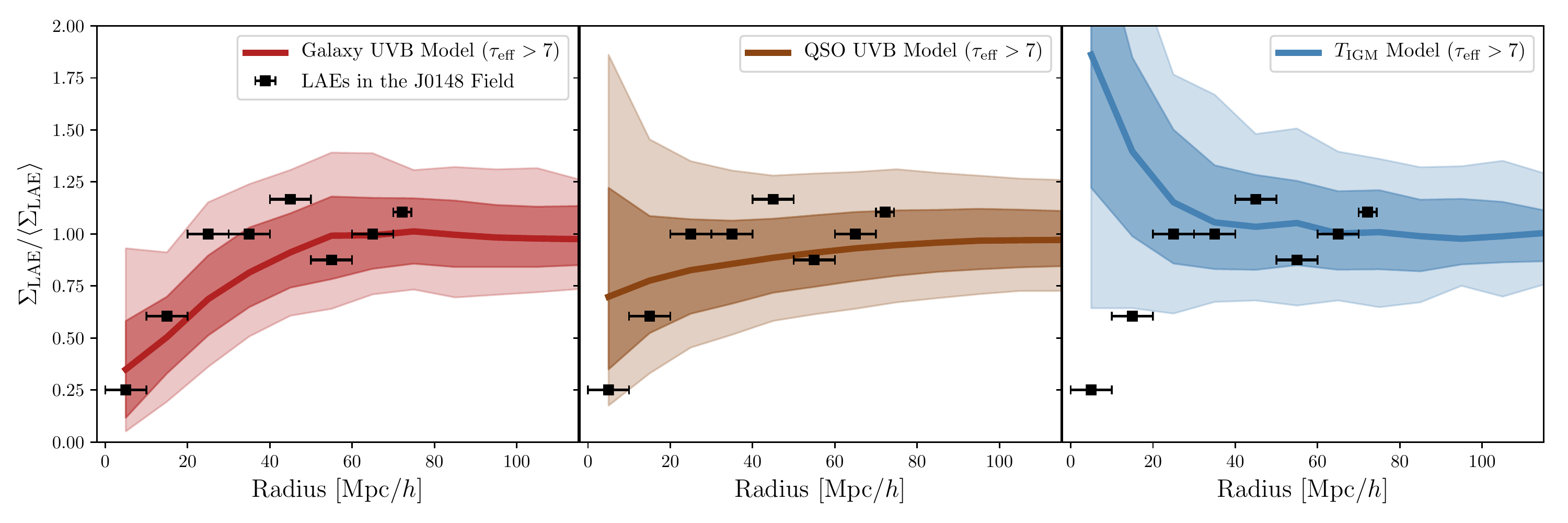}
   \caption{Comparison of the observed radial distribution of LAE candidates to model predictions.  Surface densities are plotted as a fraction of the mean.  The model predictions are for fields surrounding lines of sight with $\tau_{\rm eff} > 7$.  Predictions for the galaxy UVB (left) and temperature fluctuation (right) models are derived from \citet{davies2017}.  The QSO UVB model (center), described herein, is based on the framework of \citet{chardin2017}.  Thick lines show the mean model predictions, with 68\% and 95\% ranges from random trials shown by the dark and light shaded regions, respectively.  The data points are a single realization of this measurement and are plotted without vertical error bars.  Horizontal bars denote the radial coverage of each bin, which is 10 \hinvMpc\ for all except the outermost bin, where it is 4.5 \hinvMpc.  Model predictions are averaged over 10 \hinvMpc\ bins everywhere. \label{fig:model_compare}}
   \end{center}
   \end{minipage}
   \vspace{0.1in}
\end{figure*}

The UVB and temperature fluctuations were computed in a volume $546$ \hinvMpc\ on a side, using the semi-numerical reionization code \texttt{21cmFAST} \citep{mesinger2011} to compute the density field and dark matter halo distributions (minimum halo mass of $2\times10^9$ M$_\odot$). For the galaxy UVB model, ionizing luminosities were assigned to dark matter halos via abundance matching to the \citet{bouwens2015} (non-ionizing) UV luminosity function, allowing the ratio of ionizing to non-ionizing luminosities to be a free (but uniform) parameter that is later chosen to produce a predetermined mean \hi\ photoionization rate. The UVB was then computed on a coarse 156$^3$ grid with a spatially-fluctuating mean free path of ionizing photons following the method of \citet{davies2016}, assuming an average mean free path of $10.5$ \hinvMpc.  As noted above, this is roughly a factor of three smaller than would be expected from a naive extrapolation of lower-redshift values \citep{worseck2014}, although it is possible that some of the highest redshift ($z \sim 5$) direct measurements are biased high \citep{daloisio2018}. 

Motivated by the possibility of a QSO dominated UVB (\citealt{giallongo2015,chardin2017}, but see \citealt{mcgreer2018,parsa2018}), here we extend the modeling framework of \citet{davies2017} to include a simple QSO model of UVB fluctuations. In this model we abundance matched
\footnote{An alternative would be to randomly assign QSOs to massive halos.  This would presumably further weaken the correlation between density and \lya\ opacity seen in Figure~\ref{fig:model_compare}.}
 dark matter halos to QSOs from the \citet{giallongo2015} luminosity function and assumed the \citet{lusso2015} QSO spectrum to compute their ionizing luminosities, with zero contribution to the UVB from galaxies. We note that this is in some sense a more extreme QSO model than the one used by \citet{chardin2017}, who included a minor contribution from galaxies.  We choose an average mean free path of $42$ \hinvMpc, which is much longer than the one in our galaxy model, to reproduce the measured UVB strength at $z\sim5.7$ \citep{daloisio2018}.  As in the galaxy UVB and \citet{chardin2017} QSO models, the mean free path varies spatially, scaling with the local \hi\ ionization rate, $\Gamma_{\rm H\, I}$, and the overdensity, $\Delta = \rho/\bar{\rho}$, as $\lambda_{\rm mfp} \propto \Gamma_{\rm H\, I}^{2/3}\Delta^{-1}$. Finite lifetime effects may play a large role in the structure of the radiation field (e.g. \citealt{croft2004,davies2017a}), but for computational simplicity and consistency with \citet{chardin2017} we assumed a static QSO population.  Example slices through the galaxy UVB and QSO UVB radiation fields are shown in Figure~\ref{fig:model_maps}.

To model temperature fluctuations, we used \texttt{21cmFAST} to compute a reionization redshift field (mean $z_{\rm reion} = 8.8$; see Figure~3 in \citealt{davies2017}), and then assumed that the gas in each cell is heated to $T_{\rm reion}=30,000$~K at its corresponding reionization redshift. The subsequent thermal evolution of each cell, predominantly cooling via inverse Compton scattering off of CMB photons and through adiabatic expansion, was then integrated from its reionization redshift to $z=5.7$ following a method similar to \citet{uptonsanderbeck2016}.  The temperatures are illustrated in Figure~\ref{fig:model_maps}.

The opacity of the Ly$\alpha$ forest was computed for 1,000,000 $50$ \hinvMpc\ skewers in each model using the fluctuating Gunn-Peterson approximation \citep{weinberg1997} with optical depths re-scaled to reproduce the median effective optical depth of the Ly$\alpha$ forest on 50 \hinvMpc\ scales at $z=5.7$ ($\tau_{\rm eff}\sim3.5$). All three models, including our new QSO UVB model, generally reproduce the distribution of observed $\tau_{\rm eff}$ from \citet{becker2015} (Figure~\ref{fig:taudist}).

To predict the distribution of LAEs in these simulations, we assigned Ly$\alpha$ equivalent widths to galaxies using the probabilistic prescription presented in \citet{dijkstra2012}. Galaxies were assumed to have a fixed UV spectrum $F_\lambda\propto\lambda^{-2}$, a narrow Ly$\alpha$ emission line, and a cutoff at wavelengths shorter than rest-frame Ly$\alpha$ due to the onset of Ly$\alpha$ forest absorption. We then selected LAEs in mock narrowband observations using the same filters and color cuts as \citet{ouchi2008}. As discussed in \citet{dijkstra2012}, there is a moderate inconsistency in their approach that tends to overproduce the total number of LAEs.  To adjust for this, we discard a random $\sim50\%$ of the selected LAEs, thereby matching the normalization of the \citet{ouchi2008} LAE luminosity function.

In Figure~\ref{fig:model_compare} we compare the model predictions for the radial distribution of LAEs in fields with $\tau_{\rm eff}>7$ to our observations of the J0148 field.  Surface densities as a fraction of the mean are shown in order to minimize the impact of observational incompleteness and uncertainties in the model normalizations.  Overall we find excellent agreement with the predictions of the galaxy UVB model, with the observations closely following the expected decrease in surface density close to the QSO line of sight.  Such a strong decrease is not generally expected for the QSO model, in which UVB and density fluctuations are less tightly coupled; however, our innermost data point still falls within the 95\% expected bounds.  In contrast, the temperature model appears strongly disfavored.  Whereas an upturn in galaxy density is expected near the quasar line of sight, the opposite trend is observed.

\subsection{Caveats}\label{sec:caveats}

At face value, the results presented here support a picture wherein high IGM \lya\ opacity near $z \sim 6$ occurs in low density regions.  As discussed above, this is consistent with a model wherein large-scale opacity fluctuations are created by spatial variations in the UVB, and disfavors a scenario wherein the opacity fluctuations are due mainly to temperature variations.  While this conclusion may seem straightforward, however, it is worth considering a few caveats.

Our basic observational result is that the projected surface density of LAE candidates around the quasar line of sight is significantly lower than the ``background'' density at larger radii.  This is a self-consistent comparison and should be robust to incompleteness provided that our selection efficiency for LAE candidates is roughly constant across the field.   Given that the correction factors listed in Table~\ref{tab:results} are close to unity at all radii, this is probably close to the truth.  As in all narrow-band surveys for high-redshift LAEs, our sample will be contaminated at some level by low-redshift interlopers and other spurious sources.  \citet{Shibuya2017b} estimate a low ($\sim$8\%) contamination rate based on spectroscopic follow-up of $z=5.7$ LAE candidates selected using criteria similar to those used here, although the spectroscopically confirmed sources are all brighter than  $NB816 = 25.3$, 0.7 magnitudes brighter than our faintest sources.  So long as contaminants do not dominate our sample (which seems unlikely, given our broad agreement with the \citet{ouchi2008} number densities), contamination should not significantly affect our fundamental result.  If the contaminants are distributed randomly in the field, then removing the contaminants would tend to decrease the surface density of sources by the largest fraction in regions where the total source density is low.  This would tend to enhance the deficit of LAEs around J0148.  The contaminants are unlikely to trace the spatial distribution of genuine LAEs.  Nevertheless, we can evaluate the potential impact of contaminants in this case.  In trials where we randomly reject 30\% of all sources as interlopers, the fraction of randomly placed 20 \hinvMpc\ radius apertures that contain a number of enclosed sources less than or equal to the number of sources within 20 \hinvMpc\ of the quasar position is $\lesssim$7\% in 95\% of trials.  The deficit of LAE candidates around the quasar is therefore likely to be robust to high levels of contamination, even in this unphysical scenario.  

A second possible concern is how accurately LAEs trace the underlying density field on these scales.  In imaging of two high-redshift quasar fields, \citet{diaz2014} find that the spatial distribution of narrow band-selected LAEs at $z=5.7$ is not well matched by the distribution of broad band-selected Lyman Break Galaxy (LBG) candidates chosen to lie near $z \sim 5.7$.  \citet{ota2018} find a similar result for LAEs at $z = 6.6$ and $z > 6$ LBGs in the field of a high-redshift quasar.  Without spectroscopic redshifts of the broad-band selected objects, however, it is difficult to know how closely these populations overlap in redshift space.  Our baseline assumption is that galaxies that intrinsically emit \lya\ radiation trace the underlying density field on large ($\gtrsim$ Mpc) scales.  This should be true even if LAEs typically reside in lower-mass halos than UV-selected LBGs at the same redshift \citep[e.g.,][]{ouchi2010,bielby2016}.  

Perhaps a more serious concern is whether \lya\ emission from galaxies could be suppressed in the vicinity of a deep Gunn-Peterson trough.  After all, we selected this region because it exhibited very strong \lya\ absorption along the quasar line of sight.  it is therefore worth asking whether galaxy \lya\ emission could also be strongly extinguished.  There is strong evidence that the fraction of star-forming galaxies with detectable \lya\ emission declines at $z > 6$ \citep{fontana2010,treu2013,tilvi2014,schenker2014,pentericci2014}.  This decline is typically attributed to scattering by neutral gas along the line of sight, an indication of ongoing reionization \citep[e.g.,][]{bolton2013,mesinger2015,mason2017}.  In principle there could be remaining patches of neutral gas near $z \sim 5.7$ along the J0148 line of sight.  We note, however, that the quasar spectrum shows transmission in the \lyb\ forest, which requires a highly ionized IGM, near the central redshift of the \nb\ filter (Figure~\ref{fig:trough}).  An explanation for the lack of LAEs in this region that appeals to neutral gas in the IGM would also need to explain this transmission.  

Alternatively, a locally low UVB could enhance the apparent deficit of sources by making LAEs more difficult to detect.  If the UVB is highly suppressed, then dense circum-galactic gas may become increasingly neutral.  This would tend to scatter \lya\ out to larger galactic radii, decreasing its surface brightness \citep[e.g.,][]{sadoun2017,weinberger2018}.  In this scenario, the striking deficit of LAEs around the J0148 line of sight could be due to a lower-than-average number density of galaxies combined with \lya\ suppression.  In principle we can test this possibility by conducting a separate survey for continuum-selected galaxies.  Such a search would need to be deep enough to detect a sufficient number of sources to probe the density field, and would need photometric redshifts (or spectroscopic redshifts from lines other than \lya) that were sufficiently accurate to identify galaxies within the redshift range of the \nb\ filter and/or the full J0148 absorption trough.  As shown by \citet{davies2017}, the ability to probe the entire length of the trough means that a survey for continuum-selected galaxies could provide an even stronger test of the \taueff\ fluctuation models than the LAE results presented here.

\section{Summary}\label{sec:summary}

We have conducted a search for \lya-emitting galaxies at $z=5.7$ in the field of the $z=6.0$ quasar ULAS J0148+0600, whose spectrum contains an exceptionally long (110~\hinvMpc) and opaque ($\tau_{\rm eff} \gtrsim 7$) \lya\ absorption trough.  These observations are designed to test competing models for the origin of IGM \lya\ opacity fluctuations at $z \sim 6$ \citep{fan2006b,becker2015,bosman2018}.  If the fluctuations are due to spatial variations in an ionizing UV background dominated by galaxies, then the J0148 trough should arise in a low-density region \citep{davies2016}.  For fluctuations in a QSO-dominated UVB \citep{chardin2017}, high opacity regions can have a wide range of densities provided the local UVB is low.  If temperature fluctuations are responsible, then such troughs should trace high-density regions \citep{daloisio2015}.  

In 1.8 square degrees of Subaru Hyper Suprime-Cam imaging we identify 806 LAE candidates down to $NB816 = 26.0$ via their narrow-band excess.  The spatial distribution of these galaxies shows a clear deficit within $\sim$20~\hinvMpc\ projected distance of the quasar line of sight.  This result comes from a self-consistent comparison of the projected number density of galaxies across the HSC field, and should be minimally sensitive to completeness.  Contamination of our sample by lower-redshift and spurious sources is also unlikely to create such a large absence at the center of the field.  We are therefore confident that there is a genuine scarcity of LAEs near the quasar position.

The dearth of LAE candidates near the quasar line of sight is consistent, at face value, with models wherein the observed scatter in IGM \lya\ opacity near $z \sim 6$ is driven by spatial fluctuations in the UVB, and disfavors a model based on temperature variations.  The radial distribution of LAEs follows the expected profile for a galaxy-dominated UVB, although it is also within the broad distribution of profiles expected for a QSO UVB.  LAE surveys in additional fields surrounding deep troughs may help to determine whether a galaxy or QSO UVB is preferred.  

In the galaxy UVB model, the J0148 trough arises from a void where the ionizing radiation field is suppressed, increasing the hydrogen neutral fraction and therefore the \lya\ opacity.  The association of a high-opacity line of sight with a low-density region would be the opposite of what is expected at lower redshifts, where the hydrogen ionizing background is believed to be roughly uniform on large scales.  For example, regions of exceptionally strong \lya\ forest absorption at $z \sim 2-3$ have been associated with large-scale overdensities \citep[e.g.,][]{lee2016,cai2016}.  

If UVB fluctuations are indeed present it could have significant implications for reionization.  As noted by \citet{daloisio2018}, in order to explain the full distribution of IGM \lya\ opacities within the fluctuating UVB model, a short and spatially varying mean free path must be coupled with a mean ionizing emissivity that is a factor of $\sim$2 higher than what has been previously inferred at these redshifts.  It is unclear whether a higher emissivity tempered by a shorter mean free path would necessarily produce a more rapid reionization, but these factors are clearly important for reionization models.  More generally, any successful reionization model would have to predict large-scale UVB fluctuations at $z \lesssim 6$.

We have assumed that LAEs are reliable tracers of the density field on large scales.  It is important to consider, however, whether galaxy \lya\ emission could be scattered in regions of extreme IGM \lya\ opacity.  A survey in the same region for continuum-selected galaxies would help to determine whether there are relatively few LAEs near the J0148 line of sight because this is a genuinely low-density region or because galaxy \lya\ emission in this region is suppressed.  \lya\ scattering would need to decrease the number density of observable LAEs by a factor of $\sim$2--3 in order to fully explain the deficit around the J0148 line of sight.  Both density and scattering effects may be important, however.

Finally, we note that this only a single line of sight, and we have probed only the high-opacity end of the \taueff\ distribution.  Observations of additional fields spanning a range in line-of-sight \lya\ opacity would help to clarify which, if any, of the models considered here are correct.

\acknowledgments

We thank Anson D'Aloisio, Jonathan Chardin, and Martin Haehnelt for helpful comments.  GDB, EB, and CD are supported by the National Science Foundation through grant AST-1615814.  SRF is supported by NASA through award NNX15AK80G. SRF also acknowledges a NASA contract supporting the ÒWFIRST Extragalactic Potential Observations (EXPO) Science Investigation TeamÓ (15-WFIRST15-0004), administered by GSFC.

The Hyper Suprime-Cam (HSC) collaboration includes the astronomical communities of Japan and Taiwan, and Princeton University. The HSC instrumentation and software were developed by the National Astronomical Observatory of Japan (NAOJ), the Kavli Institute for the Physics and Mathematics of the Universe (Kavli IPMU), the University of Tokyo, the High Energy Accelerator Research Organization (KEK), the Academia Sinica Institute for Astronomy and Astrophysics in Taiwan (ASIAA), and Princeton University. Funding was contributed by the FIRST program from Japanese Cabinet Office, the Ministry of Education, Culture, Sports, Science and Technology (MEXT), the Japan Society for the Promotion of Science (JSPS), Japan Science and Technology Agency (JST), the Toray Science Foundation, NAOJ, Kavli IPMU, KEK, ASIAA, and Princeton University. 

This paper makes use of software developed for the Large Synoptic Survey Telescope. We thank the LSST Project for making their code available as free software at  http://dm.lsst.org.

The Pan-STARRS1 Surveys (PS1) have been made possible through contributions of the Institute for Astronomy, the University of Hawaii, the Pan-STARRS Project Office, the Max-Planck Society and its participating institutes, the Max Planck Institute for Astronomy, Heidelberg and the Max Planck Institute for Extraterrestrial Physics, Garching, The Johns Hopkins University, Durham University, the University of Edinburgh, QueenÕs University Belfast, the Harvard-Smithsonian Center for Astrophysics, the Las Cumbres Observatory Global Telescope Network Incorporated, the National Central University of Taiwan, the Space Telescope Science Institute, the National Aeronautics and Space Administration under Grant No. NNX08AR22G issued through the Planetary Science Division of the NASA Science Mission Directorate, the National Science Foundation under Grant No. AST-1238877, the University of Maryland, and Eotvos Lorand University (ELTE), the Los Alamos National Laboratory, and the Gordon and Betty Moore Foundation.

\bibliographystyle{aasjournal} \bibliography{j0148_laes_refs}

\end{document}